\documentclass[preprint,12pt]{elsarticle}
\usepackage{amssymb}
\usepackage{graphicx}
\usepackage{caption}
\usepackage{subcaption}
\journal{Physica D}
\begin{document}
\begin{frontmatter}
\title{A minimal model for synaptic integration in simple neurons}
\author[label1]{Adrian Alva}
\address[label1]{Center for Computational Natural Sciences and Bioinformatics, IIIT Hyderabad}
\ead{adrianjoseph.alva@research.iiit.ac.in}
\author[label1]{Harjinder Singh}
\ead{laltu@iiit.ac.in}
\begin{abstract}
Synaptic integration is a prominent aspect of neuronal information processing. The detailed mechanisms that modulate synaptic inputs determine the computational properties of any given neuron. We study a simple model for the summation of excitatory inputs from synapses and illustrate its use by characterizing some functional properties of postsynaptic neurons. In this regard, we study the response of postsynaptic neurons as defined by the model to two well known noise driven processes: stochastic and coherence resonance. The model requires a small number of parameters and is especially useful to isolate the role of integration mechanisms that rely on summation of inputs with little dendritic processing.
\end{abstract}
\end{frontmatter}
\section{Introduction}
\label{sec:intro}
Information processing in neurons is a complex phenomenon that is affected by, among other things, the kind of stimulus encoded, type of nerve cells involved, ion channels and dendritic morphology. Individual neurons are typically specialized enough to be treated as computational units, capable of performing a variety of tasks \cite{segev}. The most prominent of these is the integration of information from presynaptic inputs. Each synapse, when stimulated by the appropriate neurotransmitter, gives rise to excitatory postsynaptic potentials (EPSP) which propagate down the dendrites and are ``summed" in the cell body of the postsynaptic neuron \cite{kandeL2013,alberts2014,purves2004}. Subsequently, an action potential is generated when the membrane potential crosses a threshold. For many important classes of neurons, additional processes modulate synaptic integration. These include nonlinearities in dendritic responses to postsynaptic currents \cite{stuart}, variability in dendritic structure, inputs from inhibitory synapses and the active properties of dendrites \cite{spruston_pyramidal}. The rules of summation usually involve nonlinearities themselves in that the net activation in the cell body of the neuron may be slightly more or less than the arithmetic sum of the processed dendritic inputs \cite{poirazi,Hao21906}. Understanding how these factors affect synaptic integration requires detailed biophysical simulations incorporating several spatial and temporal scales. Nevertheless, synaptic integration in its bare minimum role as a summation of dendritic impulses can still demonstrate the functional advantages of postsynaptic neurons. In this article, we pursue such an approach and describe a minimal model of synaptic integration. The model is minimal in the sense that synaptic currents apply simple perturbations to the phase space of the postsynaptic neuron which simplifies the analysis of the resulting dynamics. Moreover, a small number of biophysically relevant parameters are used. We make use of the well known FitzHugh-Nagumo (FN) equations to simulate single neurons. The FN model is prototypical of many excitable systems  and is able to capture different modes of neuronal spiking behaviour. In his seminal work, FitzHugh \cite{fitzhugh61} showed that the phase space of the FN model could be partitioned into distinct regions, each of which corresponded to a certain physiological ``state" of the neuron such as active, refractory, etc. He also showed that the FN equations could be reduced from the famous Hodgkin-Huxley model \cite{hodgkinhuxley} which is a realistic description of neuron electrophysiology. Thus, the FN model trades detail for simplicity, yet retains the essential features of neuronal excitability.\\ 
As a first approximation to modelling excitatory postsynaptic currents, we make use of rectangular pulses. These are obtained from presynaptic membrane potentials by simple application of a Heaviside function. The Heaviside function is suitably parameterized to control the height and width of the pulse current. Each synapse thus contributes a series of rectangular pulses to the total current stimulating the postsynaptic neuron. In the next section, we briefly introduce the dynamics of the FN equations and investigate the excitation threshold for different numbers of synapses. Next, we formulate the model and explore its validity and limitations. Then, we illustrate a possible application by exploring two noise driven processes: stochastic and coherence resonance.  Finally, we highlight the scope for extending the model to incorporate other aspects of synaptic integration without compromising the simplicity of implementation. Biologically relevant observations are highlighted throughout. In starting with a minimum set of parameters and a simple model for synaptic inputs, we take a bottom-up approach that allows us to isolate the functional properties of synaptic integration made possible by summation of excitatory inputs alone (as opposed to other mechanisms like dendritic processing for instance). 
\section{Excitation threshold in the FN model}
The FN equations consist of two state variables whose dynamics operates on significantly different timescales. The fast ``voltage" variable, $v$, mimics a neuron's membrane potential and is complemented by a slow ``recovery" variable, $w$. The dynamical equations are given by:
\begin{equation}
    \epsilon\dot{v}=v(v-a)(1-v)-w+I(t)
    \label{eqn:1}
\end{equation}
\begin{equation}
    \dot{w}=v-w-b
    \label{eqn:2}
\end{equation}
where the parameters $a$ and $b$ have values $0.5$ and $0.15$ respectively and $\epsilon$ represents the ratio of the two timescales and is typically set to $0.005$. A time dependent current injection into the neuron is modelled by the function $I$. In contrast to the all-or-none nature of neuronal firing patterns, the FN model does not exhibit a well defined threshold for excitability \cite{desroches13,mitry}. That is to say, action potentials with a range of amplitudes are possible depending on the choice of the current term $I(t)$. However, there is a sense in which one can approximate a firing threshold in the FN model and this requires the analysis of a family of trajectories known as \textit{canards}  \cite{benoit,diener,izhikevich2007,durham,popovi}. We assume that the initial condition lies on the intersection of the nullclines and that the current term $I(t)$ is null at $t=0$ (Fig. 1(a)). Physiologically, this corresponds to a neuron at its resting membrane potential. If at some arbitrary time, a current is applied in the form of a rectangular step stimulus (of infinite duration), the cubic nullcline $(\dot{v}=0)$ shifts upwards and the formerly stable initial condition is now below the new fixed point. Depending on how much the nullcline shifts upward, the resulting trajectory can be either a small-amplitude excitation (Fig. 1(b)), an intermediate amplitude excitation (Fig. 1(c)), or a large amplitude excitation that resembles an action potential (Fig. 1(d)).
\begin{figure}[ht]
    \centering
    \includegraphics[width=10cm, height=8cm]{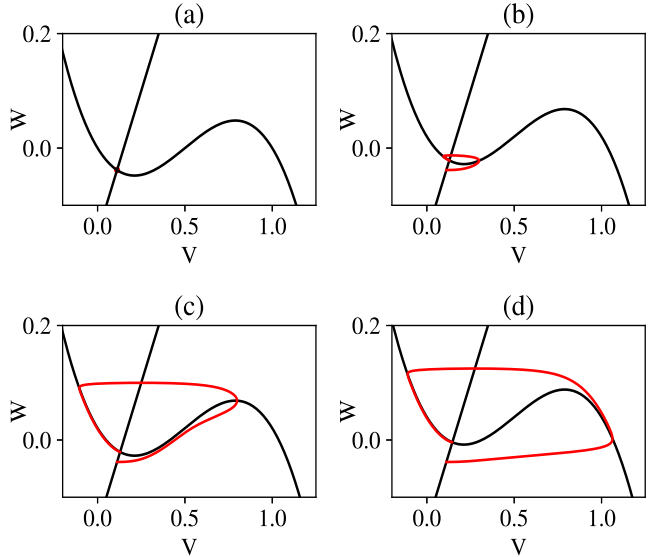}
    \caption{(a) A resting neuron with the state point on the intersection of the nullclines. (b) A small amplitude excitation generated by setting $I=0.02$ (c) An intermediate amplitude excitation ($I=0.0206662)$ that represents a canard trajectory. (d) A large amplitude excitation generated by setting $I=0.04$. Equations (1,2) were simulated with a fourth order Runge-Kutta algorithm with time step equal to $0.0001$ s. The rectangular step current was applied at $t=0.01$ s, and the trajectory was integrated for $3$ s. The initial condition $(v_{0},w_{0})\approx(0.11151, -0.03849)$ corresponds to the fixed point of the unstimulated system ($I=0$).} 
    \label{fig:1}
\end{figure}
The height of the instantaneous step current therefore determines the amplitude of the resulting excitation. Moreover, the transition from small amplitude excitations to large amplitude excitations occurs in an exponentially small ($O(e^{-\epsilon})$) region of the parameter range $I$. Within this range, trajectories known as canards track the middle branch of the cubic nullcline for a relatively long ($O(\frac{1}{\epsilon})$) time (Fig. 1(c)). The middle branch of the cubic nullcline is called the \textit{repelling slow manifold} since nearby trajectories diverge sharply from it. The right and left branches in contrast are the \textit{attracting slow manifolds} since nearby trajectories converge onto them. Informally, canards are trajectories that follow both the attracting and repelling branches of the slow manifold for a considerable amount of time \cite{izhikevich2007}. Fig. 1(c) shows a maximal canard, i.e. a trajectory that tracks the unstable branch of the cubic nullcline for the longest time. This trajectory passes through the local maximum of the cubic nullcline. The maximal canard is also called the \textit{quasithreshold separatrix} since neighbouring trajectories diverge sharply to its left or right \cite{fitzhugh61}. The quasithreshold separatrix can loosely be thought of as the neuron's threshold if we treat all trajectories to its left as subthreshold responses and trajectories to its right as action potentials. \\
\begin{figure}[t]
    \centering
    \includegraphics[width=10cm, height=8cm]{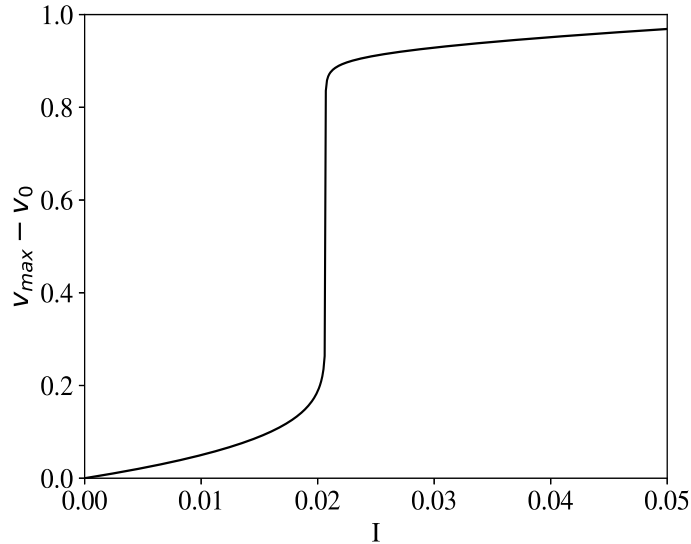}
    \caption{Stimulus-response curve for an instantaneous step current. Despite the absence of a rigid firing threshold, the canard explosion allows us to posit a firing manifold for the FN system. The step current is activated at $t=0.01$ s and the trajectories are simulated for $3$ s.}
    \label{fig: 2}
\end{figure}
If the parameter $I$ is made sufficiently large so that the fixed point lies on the middle branch of the cubic nullcline, a stable limit cycle results (via a supercritical Hopf bifurcation). Here too, for an exponentially small range of $I$ values beyond the Hopf bifurcation point, small amplitude limit cycles transition into large amplitude limit cycles in much the same way that small amplitude trajectories transitioned into large amplitude ones in Fig. 1. For the rest of this paper, we assume that the fixed point lies on the left branch of the cubic nullcline. In other words, we will only be concerned with the FN model in its excitable regime (as opposed to the limit cycle regime beyond the Hopf bifurcation point). 
\\Suppose that an instantaneous step current of infinite duration (as applied in Fig. 1) is applied to the system. We can obtain a \textit{stimulus-response} curve (Fig. 2) by plotting the voltage difference from the resting value ($v_{\rm{max}}-v_{0}$) that a given step current $I$ is able to generate. We notice a sharp transition just beyond $I=0.02$ from small to large amplitude excitations. This is sometimes referred to as a \textit{canard transition/explosion}. The sharpness of the transition allows us to approximate neuronal firings in the FN model as all or none, since intermediate amplitude excitations are sandwiched into a small threshold \textit{manifold} that is rarely ever encountered by trajectories.\\
\begin{figure}[t]
    \centering
    \includegraphics[width=10cm, height=8cm]{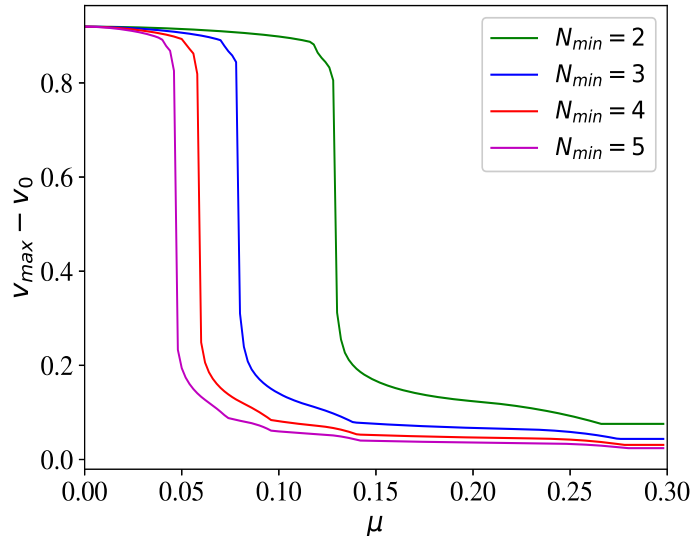}
        \caption{Threshold curves as a function of the time gap ($\mu$) between overlap events. Curves are shown for different values of $N_{\rm{min}}$. The firing threshold, $I_{\rm{T}}$, is set at $0.027$ and pulse width, $\tau=0.3$. Fourth order Runge-Kutta was used to compute the curves with step size $0.00005$.}
        \label{fig:3}
\end{figure}
Depending on the neurotransmitter and number density of its receptor, an action potential in the presynaptic neuron will lead to a small current injected into the postsynaptic dendritic membrane. This in turn generates a small amplitude excitatory postsynaptic potential (EPSP). The amplitude of a single EPSP in most cases is insufficient to elicit postsynaptic firing. This is largely due to attenuation in passive dendrites. Typical EPSP values from individual synapses range anywhere from $\frac{1}{15}$ to $\frac{1}{400}$ of the firing threshold \cite{lisman}. Figure 2 shows that there is a monotonic relationship between current amplitude and the EPSP height for \textit{subthreshold} responses (before the canard transition). Thus, if a sequence of rectangular pulses of identical heights overlap in step, at some point the current will exceed the threshold manifold. We scale the height of each current pulse such that a minimum number, $N_{\rm{min}}$, of current pulses can generate an action potential. A threshold value for firing is chosen (to the right of the canard transition in Fig. 2), denoted $I_{\rm{T}}$. The height of each current pulse is therefore $\frac{I_{\rm{T}}}{N{\rm{min}}}$. We denote the width of the pulse by $\tau$. A higher value for $\tau$ indicates a longer effective time available for a sequence of pulses to overlap and cross the firing threshold.  As in real neurons, the current pulses must overlap quickly enough to prevent the subthreshold membrane potential from falling. We can visualize this by letting $N_{\rm{min}}$ pulses each of height $\frac{I_{\rm{T}}}{N{\rm{min}}}$, overlap with an identical time gap denoted by $\mu$. Such a ``staircase current" is an idealization but the width of the staircase ($\mu$) can be thought of as the average time between successive overlaps in the real case. Figure 3 shows the corresponding threshold curves for different settings of $N_{\rm{min}}$. We notice that there is effectively a maximum time gap exceeding which, the pulse current sequence fails to evoke a postsynaptic action potential. The maximum time gap is seen to decrease as the number of synaptic inputs ($N_{\rm{min}}$) increases. This is because a large number of small height pulses apply tiny upward shifts or nudges to the cubic nullcline. Consequently, the trajectory responds by drifting towards the new fixed point (since the net current is still subthreshold). For higher $N_{\rm{min}}$, the nullcline undergoes many small magnitude, upward nudges. The smaller the nudge, the better the trajectory is able to track the shifted nullcline. This means that the component of the trajectory's drift in the $w$-direction is higher for high $N_{\rm{min}}$. As a result, by the time the final pulse arrives, the voltage of a neuron with larger $N_{\rm{min}}$ would have climbed less than one with a lower $N_{\rm{min}}$ (due to increased drift in the $w$-direction). This leads to the canard transition for higher $N_{\rm{min}}$ occurring \textit{earlier} than that for lower $N_{\rm{min}}$.\par
Low time gap requirements are especially relevant to neurons that operate as \textit{coincidence detectors} \cite{segev,kandeL2013}. The temporal resolution of coincidence detection is largely determined by parameters intrinsic to the neuron such as its membrane time constant. Figure 3 shows that even with these parameters remaining constant (no change in $a$, $b$ and $\epsilon$), modulating synaptic currents towards lower values can improve this temporal resolution.\par
Neurons are also stimulated by inhibitory, \textit{hyperpolarizing} currents called anodal currents. In some cases, it is possible for an anodal current to cause an action potential. The mechanism by which this happens is referred to as \textit{anodal break excitation} \cite{fitzhugh76}. In the FN phase space, anodal break excitation is possible because the cubic nullcline shifts downwards when a hyperpolarizing current is applied (negative values for $I(t)$). The trajectory responds by settling onto the new fixed point. When the anodal pulse terminates, the nullcline shifts upwards to its former position which can leave the state point sufficiently below the original fixed point to evoke firing. Although our focus here is on the analysis of excitatory stimuli, threshold curves analogous to Fig. 3 can be obtained for anodal currents. This is shown in Fig. 4.
\begin{figure}[t]
    \centering
    \includegraphics[width=10cm, height=8cm]{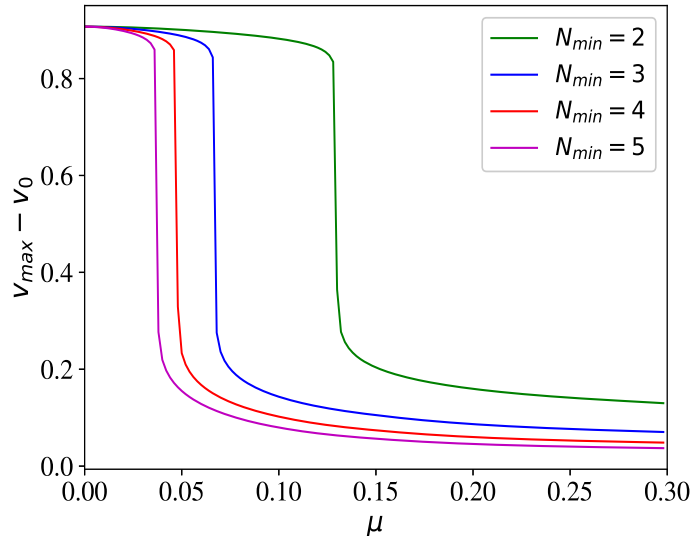}
        \caption{Threshold curves for anodal break excitation. Curves are shown for different values of $N_{\rm{min}}$. The firing threshold, $I_{\rm{T}}$, is set at $-0.05$ and pulse width, $\tau=0.3$. Fourth order Runge-Kutta was used to compute the curves with step size $0.00005$.}
        \label{fig:4}
\end{figure}
The observations are qualitatively similar to the excitatory case. There are certain features of the threshold curves in Fig. 3 that we have not explained yet. For instance, we notice that in the suprathreshold regime (low values of $\mu$), the threshold curve goes through an inverted S-shape/inflection point (just before the canard transition). This can be explained as follows. The maximum voltage ($v_{\rm{max}}$) attained by any trajectory corresponds to the point where the trajectory crosses the cubic nullcline from the left (eg. Fig 1d), since to the right of this point, the flow on the v-axis reverses direction. For low enough values of $\mu$, the trajectory hits the nullcline corresponding to the sum of \textit{all} pulses. However, for values of $\mu$ near the canard transition, the first pulse ends \textit{before} the trajectory manages to hit the final nullcline. Now, $v_{\rm{max}}$ is determined by values that range from the trajectory's intersection with the \textit{penultimate nullcline}, to the trajectory's intersection with the final nullcline. This is what causes the small inflection point. In the subthreshold regime (beyond the canard transition), we notice points where the monotonic decrease of $v_{\rm{max}}$ is non-smooth. This can be explained similarly: $v_{\rm{max}}$ is determined by the penultimate and lower nullclines. In other words, the trajectory reverses direction after crossing the nullcline of the $n\textsuperscript{th}$ pulse and traverses far enough that all pulses $>n$ fail to get it past its previously attained maximum voltage.\par
Synaptic integration of inputs exhibits two modes known as spatial and temporal summation \cite{kandeL2013,alberts2014,purves2004}. Spatial summation involves the simultaneous overlap of two or more dendritic inputs and a subsequent firing. In temporal summation, high frequency stimuli from the same dendritic input can summate to evoke firing. This requires that the time lag between stimuli be much smaller than the characteristic decay time of the postsynaptic potential. We remark here that the model we use can only exhibit spatial summation. This stems directly from the dynamics of the FN system, which belongs to a class of neuronal models termed \textit{resonators} \cite{izhikevich2007,izhikevichIJBC}. Real neurons can temporally integrate subthreshold excitatory currents if these currents arrive with sufficiently low time gaps. In our model though, even if two or more subthreshold pulses follow each other with zero time lag, then that simply corresponds to a single subthreshold pulse of finite duration which anyway cannot cause firing. 
\section{Noisy FN model for synaptic input}
\label{sec:model}
We shall treat voltages obtained from FN trajectories as action potentials whenever a threshold value, $v_{\rm{T}}$, is crossed from below. We choose $v_{\rm{T}}=0.9$ which is close to the action potential peak. Because we include noise in the presynaptic FN equations, it is possible that during the course of an action potential $v$ may cross $0.9$ several times by virtue of stochastic fluctuations. To avoid such overcounting, we specify that a low voltage ($v=0.2$) be crossed from below in between two crossings of $v_{\rm{T}}$.\par
Equation (1) is modified to include an additive noise term as follows:
\begin{equation}
    \epsilon\dot{v}=v(v-a)(1-v)-w+I(t)+\sigma\xi(t)
\end{equation}
where $\xi(t)$ is white noise corresponding to a normalized Wiener process with autocorrelation $\langle \xi(t)\xi(t')\rangle=\delta(t-t')$, and $\sigma$ (which we shall call the noise intensity) is the standard deviation of the Wiener increments. Equation (2) remains unchanged. How the presynaptic neurons themselves are stimulated depends on the type of phenomenon studied. For instance, we can let the presynaptic neurons be driven by a sinusoidal current stimulus as follows:
\begin{equation}
    I_{\rm{pre}}(t)=A\sin(2\pi ft)
\end{equation}
Denote the firing times of the $i^{\rm{th}}$ neuron by the set $\{t_{\rm{j}}\}$. The sequence of current pulses that the $i^{\rm{th}}$ synapse generates is therefore
\begin{equation}
    S_{\rm{i}}=\sum_{j}\{H(t-t_{\rm{j}})-H(t-(t_{\rm{j}}+\tau))\}
\end{equation}
where $\tau$ is the pulse width and $H$ denotes the Heaviside function. The net stimulus from all synapses is then
\begin{equation}
    I_{\rm{post}}(t)=\frac{I_{\rm{T}}}{N_{\rm{min}}}\sum_{\rm{i=1}}^{\rm{N}}S_{\rm{i}}
\end{equation}
where $I_{T}$ is the firing threshold described in section 2, $N$ is the total number of synapses and $N_{\rm{min}}\in\mathbb{N}$ is the minimum number of presynaptic firings requires to elicit a postsynaptic action potential. In other words, if $N_{\rm{min}}$ pulses overlap in close succession, then the total postsynaptic current equals $I_{\rm{T}}$ which evokes firing. \textit{Postsynaptic neurons are kept noiseless for the simulations in this paper} (i.e., $\sigma=0$). We choose this because as stated in the introduction, we are concerned with synaptic integration in its minimum role as a summation of excitatory stimuli. This will however require us to state assumptions on the \textit{source} of noise that justify our choice of keeping postsynaptic neurons noiseless. For instance, we may choose to study synaptic integration of inputs from sensory neurons whose stimuli are affected by \textit{environmental} noise. This would then justify keeping the corresponding postsynaptic neuron noiseless as it has access only to its presynaptic neurons and not the external environment.\par
In a sequence of input pulses overlapping to generate an action potential, there is always a time lag between one pulse and the next. These time lags cause a net drift of the trajectory in the leftward region of the quasithreshold separatrix of the nullcline set by the final pulse. The final pulse pushes it to the right of the separatrix and causes firing. While setting $I_{\rm{T}}$, we must avoid values close to the canard transition in Fig. 2. It is possible that for such values of $I_{\rm{T}}$, even if $N_{\rm{min}}$ pulses overlap in very close succession, the net drift can still bring the trajectory to the left of the quasithreshold separatrix of the final nullcline and prevent firing.\par
The stochastic differential equations corresponding to Equations(3,2) respectively are:
\begin{equation}
    dv=\frac{1}{\epsilon}[v(v-a)(1-v)-w+I]dt+\frac{\sigma}{\epsilon}dW
\end{equation}
\begin{equation}
    dw=(v-w-b)dt
\end{equation}
The numerical integration of Equation(7) requires a choice of interpretation of the stochastic integral used to integrate the noise term $\frac{\sigma}{\epsilon}dW$ (called the It\^{o} and Stratonovich interpretations). For additive white noise however, both interpretations evaluate to the same stochastic integral so this does not affect the choice of which algorithm we use. We choose the simple first order Euler-Maruyama scheme \cite{kloeden92,tuckwell} which gives:
\begin{equation}
    v_{n+1}=v_n+\frac{1}{\epsilon}[v_n(v_n-a)(1-v_n)-w_n+I]\delta t+\frac{\sigma}{\epsilon}\sqrt{\delta t}\Delta W_n
\end{equation}
\begin{equation}
    w_{n+1}=w_n+(v_n-w_n-b)\delta t
\end{equation}
where $\Delta W_n$ is sampled from a Gaussian distribution with unit variance and $\delta t=0.001$. Because the Euler-Maruyama method is stiff, interspike interval histograms of neuronal firings were compared with those obtained using a second order stochastic Runge-Kutta method \cite{kloeden_pearson_1977} for typical noise intensities used in this paper. The same was repeated for step size an order of magnitude lower. No significant differences were found. Interspike interval histograms (normalized) are a good approximation to the probability distribution functions of waiting times between firings. For obtaining postsynaptic trajectories, we use a standard deterministic Runge-Kutta method ($4^{\rm{th}}$ order) with step size $0.001$ (since they are kept noiseless). Sample trajectories are shown in Fig. 5 for a postsynaptic neuron stimulated by $5$ synapses.\par 
\begin{figure}[t]
    \centering
    \includegraphics[width=13cm, height=9cm]{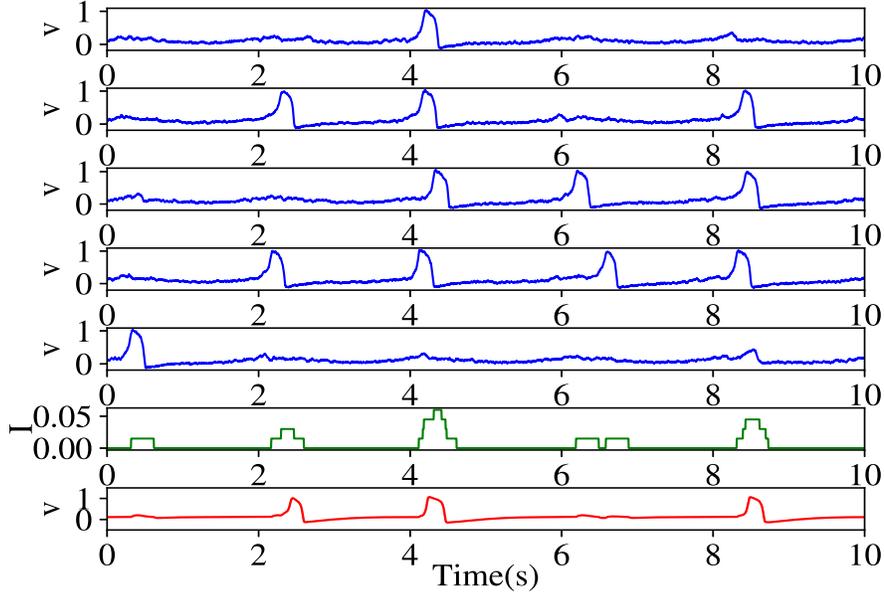}
    \caption{Voltage trajectories for $5$ presynaptic neurons (blue) and a postsynaptic neuron (red, final trace) for which $N_{\rm{min}}$=2 and $I_{\rm{T}}=0.027$. The driving frequency for all presynaptic neurons is set at $f=0.5$ Hz and subthreshold amplitude at $A=0.03$ (so that the signal by itself cannot evoke firing). The standard deviation of noise for all presynaptic neurons, $\sigma=0.001$. The net stimulus current, $I_{\rm{post}}(t)$, is shown on the second trace from below (pulse width $\tau$ is set at $0.3$.)}
    \label{fig:5}
\end{figure}
There are two caveats with defining the postsynaptic input function as in Equation(6). Suppose we choose a relatively high value for $I_{\rm{T}}$, say $0.08$ and that $N_{\rm{min}}=2$. Now if a single pulse alone stimulates the postsynaptic neuron, then $I_{\rm{post}}(t)$ will be $\frac{I_{\rm{T}}}{2}=0.04$ which is suprathreshold (see Fig. 2). Thus, a single pulse will cause firing when it is not supposed to. The second problem arises when stimulating the neuron with high number of inputs, $N$. Suppose $N=100$, $N_{\rm{min}}=5$ and that $I_{\rm{T}}=0.03$. If $50$ inputs undergo spatial summation, then $I_{\rm{post}}(t)=\frac{50*0.03}{5}=0.3$ which is well beyond the Hopf bifurcation point for $I$ (leading to limit cycles and further on, into an unstable node). Thus, large values of $N$ and small values of $N_{\rm{min}}$ can lead to bifurcations in the dynamics. This is not necessarily undesirable from a modelling perspective. For example, a Hopf bifurcation caused by large $N$ and small $N_{\rm{min}}$ might help model interesting features such as bursting dynamics \cite{lisman}. We will however not deal with such a scenario here and assume that the neuron is always in its excitable regime (where the fixed point lies on the left branch of the cubic nullcline). It is possible to artificially correct for high values of $I_{\rm{post}}(t)$ by composing it with another Heaviside function. However, we avoid this and simply point towards the validity of the model by outlining the constraints on the parameters $I_{\rm{T}}$, $N$ and $N_{\rm{min}}$:
First, we would like to prevent any number of presynaptic inputs less than $N_{\rm{min}}$ from exciting the postsynaptic neuron. The worst case corresponding to this is when $N_{\rm{min}}-1$ inputs fire (nearly) simultaneously. Then, $I_{\rm{post}}(t)=\frac{(N_{\rm{min}}-1)I_{\rm{T}}}{N_{\rm{min}}}$ which we would like to limit to some subthreshold value $I_{\rm{sub}}$(chosen appropriately from Fig. 2). Thus, we have 
\begin{equation}
    \frac{(N_{\rm{min}}-1)I_{\rm{T}}}{N_{\rm{min}}}<I_{\rm{sub}}
\end{equation}
If bifurcations are to be strictly avoided, then we limit the maximum value of $I_{\rm{post}}(t)$ to the value $I_{\rm{Hopf}}$ where the Hopf-bifurcation occurs ($I_{\rm{Hopf}}\approx0.11$):
\begin{equation}
    \frac{NI_{\rm{T}}}{N_{\rm{min}}}<I_{\rm{Hopf}}
\end{equation}
This constraint can be relaxed if such high values for postsynaptic currents are rare, or if limit cycle spikes are desired. We have said that $N_{\rm{min}}$ is a positive integer. There is no particular reason why $N_{\rm{min}}$ should be an integer except that doing so would provide a natural interpretation to the current term for postsynaptic neurons. The value of $N_{\rm{min}}$ along with $I_{\rm{T}}$ and the summation term in Eq. (6) determines how much the cubic nullcline shifts upward and so a non-integer value for $N_{\rm{min}}$ should also work. However, we will use integer values for $N_{\rm{min}}$ throughout. In the biological context, constraints (11-12) might correspond to simple neurons such as cerebellar granule cells which on average receive only 4 excitatory inputs from short dendrites \cite{spruston2009}.\par 
In the following section, we explore the effects of excitatory stimuli only. However, by the simple addition of a term to equation (5), we can incorporate the effects of inhibitory synapses. For a neuron with $N_{\rm{1}}$ excitatory inputs and $N_{\rm{2}}$ inhibitory inputs, the postsynaptic current will be as follows:
\begin{equation}
    I_{\rm{post}}(t)=\frac{I_{\rm{1}}}{n_{\rm{1}}}\sum_{i=1}^{N_{\rm{1}}} S_{\rm{i}}^{+}-\frac{I_{\rm{2}}}{n_{\rm{2}}}\sum_{i=1}^{N_{\rm{2}}} S_{\rm{i}}^{-}
\end{equation}
where $S_{\rm{i}}^{+}$ and $S_{\rm{i}}^{-}$ are the sequence of Heaviside pulses from excitatory and inhibitory synapses respectively, $n_{\rm{1}}$ and $n_{\rm{2}}$ are the minimum number of presynaptic firings required to elicit a postsynaptic action potential, when excitatory or inhibitory inputs are considered alone. The parameters $I_{\rm{1}}$, $I_{\rm{2}}$ are the corresponding firing thresholds. How to set the inhibitory firing threshold, $I_{\rm{2}}$, depends on the kind of phenomena we would like to investigate. For example, if we want to explore firings generated by inhibitory synapses via anodal break excitation, then $I_{\rm{2}}$ can be chosen appropriately from Fig. 4. On the other hand, if we are concerned only with the role of inhibiting/ hyperpolarizing currents, then $I_{\rm{2}}/n_{\rm{2}}$ can be replaced with a single parameter that describes the height of each inhibiting pulse.\par 
To illustrate some use cases of the model, we briefly investigate two well known noise driven phenomena: stochastic and coherence resonance. 
We compare presynaptic and postsynaptic responses in these effects and identify postsynaptic features that may be biologically significant.
\section{Postsynaptic responses to noisy presynaptic neurons}
\subsection{Stochastic resonance}
Noise is a ubiquitous presence in neurons and can play a significant role in certain sensory modalities. The functional role of a variety of stochastic fluctuations in interspike variability has been documented \cite{stein,faisal,mcdonnell}. The FN model has been shown to account for several noise driven phenomena such as stochastic resonance \cite{longtin, collins, nozaki}, coherence resonance \cite{lindner,pikovsky}, resonant activation and noise enhanced stability \cite{valenti}. Multiple stochastic FN neurons can be coupled to produce noise-induced synchronization \cite{acebron} and phase locking \cite{davidson}.\par
Experimental recordings of sensory neurons are often visualized in terms of interspike interval histograms (ISIH). Typical ISIH recordings have a multimodal shape with peaks at integer multiples of the external stimulus' time period. This indicates that firing occurs during a preferred phase of the external stimulus and that sometimes, a random number of cycles are skipped between successive firings \cite{longtin}. This pattern of firing is especially interesting due to its relevance to the well known phenomenon of \textit{stochastic resonance}. A classic result in the theory of stochastic processes is that noise can increase the synchronization between the response of some nonlinear systems and an external periodic signal. When this happens, the system is said to undergo stochastic resonance (SR) \cite{gammaitoni}. A simple example of the SR effect is an overdamped Brownian particle in a double well potential \cite{mcnamara}. Modulating the potential with a weak periodic signal alternately raises and lowers the two wells, although the signal by itself is insufficient to cause inter-well transitions. Adding white noise to the external forcing however enables the particle to switch wells. Intuitively, this is understood as the result of a time-scale matching condition \cite{gammaitoni}: the average waiting/residence time of the particle between transitions is comparable with half the period of the driving signal. Consequently, for a given value of the driving signal’s frequency, there exists a corresponding noise intensity that achieves the highest tuning between noise induced transitions and the driving signal.\par 
SR has been documented extensively in experimental systems, notably in sensory neurons \cite{moss,russel}. Moreover, residence-time distributions of periodically driven bistable systems with noise resemble the ISIH obtained from experimental recordings. This is consistent with the fact that neuron models like FN can be approximated as bistable systems. For the FN model in particular, the voltage equation (Eq. 3) can be cast into a form equivalent to Langevin dynamics in a one-dimensional double well potential \cite{collins}. This approximation works essentially due to the large separation of timescales set by $\epsilon \ll 1$ so that $\dot{w} \sim 0$ on the fast (nearly straight line) branch of the excursion along $v$. A noise induced "escape" from the fixed point is analogous to a switching between wells. The trajectory returns to the fixed point from the other well via a different degree of freedom (i.e., $w$) \cite{collins}.\par 
The SR effect can make detection of a weak external stimulus possible, but for the stimulus to be reliably encoded and transmitted further requires some sort of amplification mechanism.
\begin{figure}[t]
    \centering
    \includegraphics[width=9cm,height=7cm]{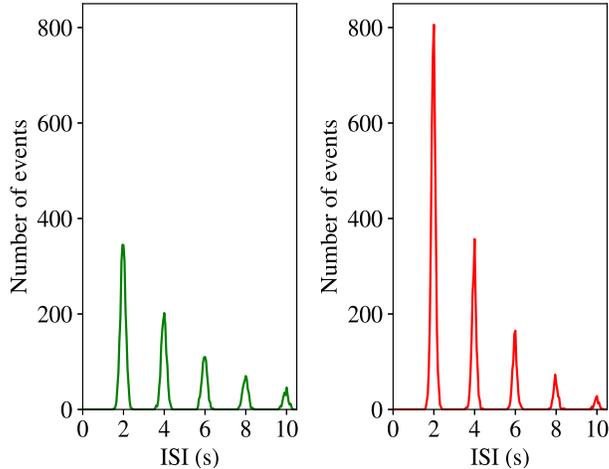}
    \caption{ISIH computed for a presynaptic (left) and postsynaptic (right) neurons. Stimulus parameters used are $A=0.03$ (subthreshold) and $f=0.5$ Hz. The noise intensity is set at $\sigma=0.001$. Both simulations were integrated for $25000$ s with step size $0.001$ s. The postsynaptic neuron is stimulated by $8$ synapses with $\tau=0.6$ s (pulse width), $N_{\rm{min}}=3$ (minimum presynaptic firings) and $I_{\rm{T}}=0.027$ (firing threshold). The bin width of the histogram is set at $0.05$ s.}
    \label{fig:6}
\end{figure}
Clearly, postsynaptic neurons play a significant role in this. In Fig. 6, we compare the ISIH obtained from a presynaptic neuron and a postsynaptic neuron with $8$ synapses. The signal chosen is subthreshold, i.e. unable to cause firing in the absence of noise. In both cases, the dominant peak occurs at the stimulus time period, $T_{\rm{0}}=2.0$ s. The picture is qualitatively the same if we choose to plot the probability density by normalizing the ISIH (peaks become rescaled then). The contribution of the dominant peak has implications in the reliability of information transmission. For instance, the area under the peak may correspond to how much information about the stimulus' period is conveyed \cite{longtin}. If this interpretation is followed, then Fig 6 shows that it is possible for subthreshold signals to be conveyed with increased reliability. Notwithstanding the encoding mechanism, we remark here that an important application of the model would be to quantify postsynaptic firing statistics using information-theoretic measures \cite{borst}. This makes quantitative, the reliability of postsynaptic spikes to encode weak stimuli rendered detectable by noise. \par 
There are different measures used to quantify stochastic resonance. Typically, the signal to noise ratio (SNR) is computed from the system response's power spectrum and plotted as a function of noise intensity \cite{gammaitoni}. A peak in the SNR plot represents the point of maximum resonance with the signal. For neuronal firings, the power spectrum of the point process corresponding to firings is typically used. Other measures include the power spectral density of the response at the signal frequency, or even the magnitude of the corresponding Fourier coefficient. We plot instead, the peak of the ISIH corresponding to the stimulus period, $T_{\rm{0}}$, as in \cite{longtin}. ISIH plots are an intuitive way to visualize how the firing statistics are distributed across intervals relevant to both the external stimulus and dynamics intrinsic to the neuron. Although ISIH peaks may be less useful for quantitative arguments than measures like the SNR, they are a good starting point and easily interpretable.
\begin{figure}[t]
    \centering
    \includegraphics[width=10cm,height=8cm]{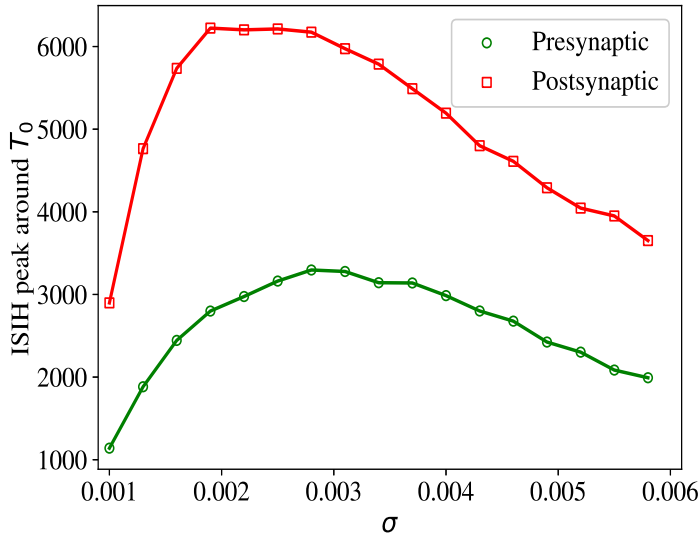}
    \caption{Stochastic resonance in pre and postsynaptic neurons. The stimulus parameters are $A=0.03$ and $f=0.5$ Hz. The postsynaptic neuron is stimulated by $5$ synapses with $N_{\rm{min}}=2$, $\tau=0.3$ s and $I_{\rm{T}}=0.035$. For both pre and postsynaptic neurons, $40$ runs, each lasting $2000$ s were used to obtain the interspike intervals. The histogram bin width was set to $0.05$ s and the peak was chosen from $0.35$ s on either side of $T_{\rm{0}}$.}
    \label{fig:7}
\end{figure}
As indicated in section 3, we are required to indicate what kind of system corresponds to our results in order to justify keeping the postsynaptic neuron noiseless. A possible application would be sensory neurons which are stimulated by an external signal and then relay their firings onto a postsynaptic neuron. Noise is then assumed to originate externally in the environment. Figure 7 shows a comparison of the stochastic resonance effect for such a system. The curve for the postsynaptic case rests above its presynaptic counterpart. The significantly higher postsynaptic ISIH peaks, also grow faster for low noise intensities. This suggests a dual functionality for postsynaptic neurons conveying sensory information (eg. interneurons from sensory afferents): They amplify the information in the intervals corresponding to $T_{\rm{0}}$, and they could be more sensitive to an increase in noise, within the low noise range. The former ensures robustness against a presynaptic neuron failing to fire. The latter is an issue that requires further investigation.\par 
Because the ISIH is multimodal (consisting of several peaks), it is possible to demonstrate stochastic resonance using the second or third peak as well (corresponding to $2T_{\rm{0}}$ and $3T_{\rm{0}}$ respectively). This becomes necessary if the first peak occurs at a noise intensity that is too high: In such a case, the condition that the neuron fires at a preferred phase is no longer valid and noise induced firings take place throughout the rising phase of the stimulus cycle \cite{longtin}. For Fig. 7 however, the stochastic resonance peaks for both pre and postsynaptic neurons were found to occur at noise intensities at which the dominant contribution to the interspike intervals is still from the stimulus period, $T_{0}$. There is also a small secondary peak at $<T_{0}$ that corresponds to limit cycle spikes when the neurons are driven above their Hopf bifurcation point.\par  
The postsynaptic parameters $N$, $N_{\rm{min}}$ and $\tau$ affect the ISIH peak at the stimulus time period, $T_{\rm{0}}$. Increasing $N$ or $\tau$ (all else being constant) increases the peak around $T_{\rm{0}}$ because of increased stimulation or increased effective time available to input pulses for overlapping. Decreasing $N_{\rm{min}}$ also increases the peak because less number of stimuli are now required to cause firing.  
\subsection{Coherence resonance}
If we let the FN system be driven purely by a noise term (so that $I_{\rm{pre}}(t)=0$), then the system can still tune to an optimal noise intensity at which the firings are most coherent. That is, at the optimal noise intensity, the characteristic correlation time of the sequence of firings is maximum (rate of decay of correlations is slowest). This effect is called coherence resonance (CR) \cite{lindner,pikovsky}. CR has been analysed and predicted by the FN model using the standard Langevin approximation for the state variable to which noise is added \cite{pikovsky}. Intuitively, the CR effect can be traced to the existence of two different characteristic times in the FN system: the activation time $t_a$ (time required to elicit firing) and the excursion time $t_e$ (duration of action potential). Depending on the noise intensity, one of these two times dominates the average pulse duration (defined as the interspike interval $=t_{a}+t_{e}$). For small values of noise intensity, $t_a \gg t_e$ and for large values of noise intensity, $t_a \ll t_e$. Coherence resonance occurs when the noise intensity is large enough ($t_a \ll t_e$) so that excitations are frequent while at the same time fluctuations in the excursion time are low (high coherence). 
To quantify CR, the coefficient of variation (CV) of the interspike intervals is typically used. That is, we first compute the sequence of firings $\{t_i\}$ of a long spike train and from it, the interspike intervals $\{T_i\}$ where $T_{i}=t_{i}-t_{i-1}$. The coefficient of variation is defined as:
\begin{equation}
    CV=\frac{\sqrt{\langle T^{2}\rangle-\langle T \rangle ^2}}{\langle T \rangle}
\end{equation}
The CV is a measure of the normalized fluctuations of the interspike intervals. Coherence resonance manifests as a minimum in the graph of CV versus noise intensity. Figure 8 plots CV versus noise intensity for a presynaptic neuron and a postsynaptic neuron with $5$ synapses.
\begin{figure}[t]
    \centering
    \includegraphics[width=10cm,height=8cm]{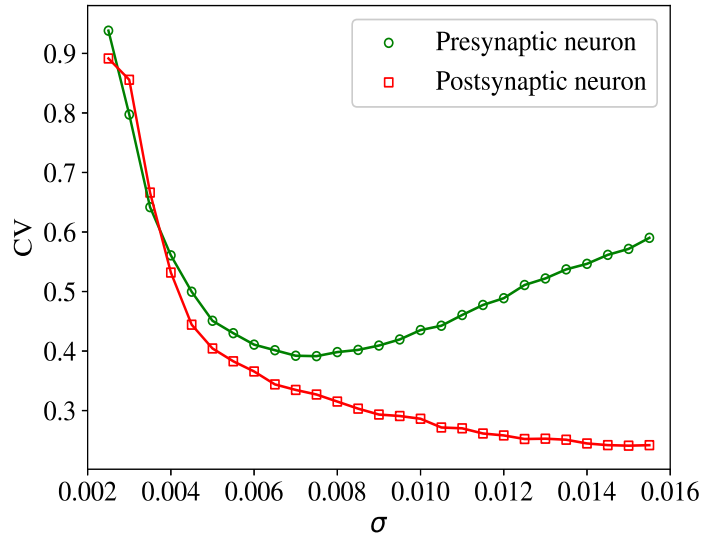}
    \caption{CV versus $\sigma$ for pre and postsynaptic neurons. The postsynaptic neuron is stimulated by $5$ synapses, with $N_{\rm{min}}=2$, $I_{\rm{T}}=0.035$ and $\tau=0.5$ s. For both pre and postsynaptic neurons, $10$ runs, each lasting $1000$ s were used to obtain the interspike intervals.}
    \label{fig:8}
\end{figure}
The presynaptic plot shows the expected shape \cite{pikovsky} whereas the postsynaptic plot deviates significantly from it as the noise intensity increases. The explanation is as follows: For presynaptic neurons at high noise intensities, CV is determined by fluctuations of the characteristic \textit{excursion time}, $t_e$ (since $t_a$ is very low). In particular, $CV\sim\sigma\langle t_e\rangle^{1/2}$ \cite{pikovsky}. Since $\langle t_e\rangle$ depends weakly on $\sigma$ for high noise intensities, CV increases linearly with $\sigma$ for presynaptic neurons as observed in Fig. 8. For postsynaptic neurons however, the excursion time is effectively constant as they are kept noiseless (i.e., the action potential of a postsynaptic neuron is unperturbed by noise). The CV is therefore determined purely by fluctuations in the \textit{activation time}, $t_a$ of the postsynaptic neuron. This is very low since the presynaptic neurons themselves exhibit low fluctuations of $t_a$ and they independently stimulate the postsynaptic neuron. As a result, CV is nearly flat at high noise. Care must be taken while using high noise intensities: In such cases, the dynamics is almost entirely dominated by noise and action potentials become indistinguishable. As a result, the spike detection procedure outlined at the beginning of section 3 can miss rapid spikes that do not reach $v=0.2$ in between. For Fig. 8, this starts to affect the CV after approximately $\sigma=0.14$.\par
Theoretically the use of high noise is unproblematic. However, we are required to state assumptions of where the noise possibly comes from to ensure biological interpretability. For the coherence resonance scenario, we assume the noise to be external (making the whole setup analogous to the sensory system assumed in section 4.1). Other interpretations may be possible but require justification for not only keeping the postsynaptic neuron noiseless but also the use of white noise for presynaptic neurons (so far we have assumed the noise to be external and therefore environmental white noise is a reasonable approximation).\par
The more coherent postsynaptic response in Fig. 8 at high noise could be relevant to neurophysiological mechanisms for amplifying order in spontaneously (random) firing networks.
\section{Discussion and Conclusions}
We have investigated a simple model for synaptic integration using rectangular pulses. The model is minimal since the phase space of the FN system is perturbed by a finite sequence of instantaneous changes (via rectangular pulses) as opposed to continuous changes in parameter values. We have illustrated the use of the model in probing noise assisted/noise driven phenomena where the noise source is assumed to be external to the system and relatively larger than intrinsic sources of noise such as channel noise and synaptic noise \cite{faisal}. A natural extension would be to let the postsynaptic responses be affected by noise as well. This is particularly relevant to internal neurological systems such as cortical networks. Channel noise for the postsynaptic case can be simulated with nonzero $\sigma$ in Eq. 3. Synaptic noise can be implemented by using an additional noise term to perturb the pulse width, $\tau$. This lets the effective time gap for spatial summation fluctuate, mimicking the stochastic opening and closing of ion channels.\par
The neurocomputational properties of a model depend on the types of bifurcations it undergoes \cite{izhikevich2007,izhikevichIJBC}. The FN model is capable of exhibiting interesting features such as tonic spiking, phasic spiking, class-1 excitability, rebound spikes, accommodation, etc \cite{izhikevichIEEE}. A further point of inquiry would be to see how postsynaptic responses vary for each of these properties as a function of parameters $N$, $N_{\rm{min}}$, $\tau$. For neurons stimulated by suprathreshold signals, different $m:n$ phase locking patterns are possible (where $m$ spikes occur every $n$ cycles) \cite{guevara}. It would be interesting to see how postsynaptic neurons respond to these deterministic phase locking patterns of presynaptic neurons.\par
For the stochastic resonance simulations, we have used signals that are in phase with each other. This is under the assumption that the presynaptic (sensory) neurons are spatially close so that they are effectively stimulated by identical signals. We included phase differences so that $I_{\rm{pre}}(t)=A\sin(2\pi ft+\phi)$ where $\phi$ is randomly sampled and independently set for the presynaptic neurons. If $\langle\phi\rangle$ is not small, the ISIH peaks of the postsynaptic response was found to decrease. Thus, anatomical features become relevant for the ability to exploit background noise if the neurons are separated enough.\par
We have used uncorrelated, white noise for our simulations. It has been shown both with experiment and analysis that coloured noise such as $1/f$ and $1/f^{2}$ noise can exhibit stochastic resonance and moreover enhance it \cite{douglas,nozaki}. Implementing this to the postsynaptic case is an important extension of the results shown here.\par 
It is known that single spikes produce unreliable synaptic transmission, with the probability of transmission ranging from $<0.1$ to $>0.9$ \cite{lisman}. This suggests the functional importance of bursting patterns which can enhance transmission. A neural code that involves presynaptic neurons firing coincident bursts is thus likely to be robust. Bursts can be simulated in the model by adding a constant term to $I(t)$. This places the fixed point of the FN system closer to the Hopf bifurcation point. For stimulus frequencies that are smaller than the frequency of the limit cycle, we can have several spikes per stimulus cycle, mimicking the bursting dynamics. Note that $I_{\rm{T}}$ will have to be obtained from the threshold plot computed by including the constant term added to $I(t)$. Probabilistic synaptic transmission is incorporated by introducing a random variable $X\in(0,1]$: Spikes are generated when the neurons cross the firing threshold $I_{\rm{T}}$ \textit{and} when $X(t)<p$ (uniform sampling) where $p$ is the probability of transmission.  





\end{document}